\documentclass[a4paper]{jpconf}
\usepackage{graphicx}
\begin{document}
\title{Spacetime Symmetries 
and $Z_3$-graded Quark Algebra}

\author{Richard Kerner}

\address{Laboratoire de Physique Th\'eorique de la Mati\`ere Condens\'ee, \\
Universit\'e Pierre et Marie Curie (Paris-VI) - CNRS UMR 7600 \\
Tour 23-13, B.C. 121, 4 Place Jussieu, 75005 Paris, France}

\ead{richard.kerner@upmc.fr}

\begin{abstract}
We investigate certain $Z_3$-graded associative algebras with cubic $Z_3$-invariant constitutive relations.
The invariant forms on finite algebras of this type are given in the low dimensional cases with two or
three generators. 

We show how the Lorentz symmetry represented by the $SL(2, {\bf C})$ group emerges naturally without any notion
of Minkowskian metric, just as the invariance group of the $Z_3$-graded cubic algebra and
its constitutive relations. Its representation is found in terms of Pauli matrices. The relationship
of this construction with the operators defining quark states is also considered, and a third-order analogue
of the Klein-Gordon equation is introduced. Cubic products of its solutions may provide the basis
for the familiar wave functions satisfying Dirac and Klein-Gordon equations.

\end{abstract}

\section{Introduction}

The Lorentz and Poincar\'e groups were established as symmetries of
the observable macroscopic world. More precisely, they were conceived in order to take into account 
the relations between the components of electromagnetic and magnetic fields as seen by different 
inertial observers from different Galilean reference frames.

Only later did Einstein extend the Lorentz transformations to space
and time coordinates, modifying also Newtonian mechanics
so that it could become invariant under the Lorentz instead of the Galilei group. In most
of the textbooks introducing the Lorentz-Poincar\'e group the accent is put on
the transformation properties of space and time coordinates, and the invariance of the
Minkowskian metric tensor $g_{\mu \nu} = diag (+,-,-,-)$. But neither the components of
$g_{\mu \nu}$, nor the space-time coordinates of an observed event can be given an intrinsic
physical meaning; they are not related to any conserved or directly observable quantities.

A more reliable physical content of Lorentz transformations is revealed when they are applied
to the observable and measurable quantities such as electric charges and currents, or frequencies
 and wavelengths of electromagnetic waves. The Lorentz transformations apply directly to the
four-current $j^{\mu} = [\rho c, {\bf j} ]$ and to the four-vector $k^{\mu} = [\omega/c, {\bf k}]$.
Two galilean observers comparing these quantities will arrive at the transformation property
which is in agreement with charge conservation and the (relativistic) Doppler effect: in the
appropriate units, one has
\begin{equation}
j^{\mu'} = \Lambda^{\mu'}_{\nu} \, j^{\nu}, \; \; \; k^{\mu'} = \Lambda^{\mu'}_{\nu} \, k^{\nu}, 
\label{reldoppler}
\end{equation}
The differential form of the Lorentz force,
\begin{equation}
\frac{d {\bf p}}{dt} = q \, {\bf E} + q \, \frac{\bf v}{c} \wedge {\bf B}
\label{Lforce}
\end{equation}
combined with the energy conservation of a charged particle
under the influence of electromagnetic field
\begin{equation}
\frac{d {\cal{E}}}{dt} = q \, {\bf E} \cdot {\bf v}
\label{EnergyL}
\end{equation}
 is also Lorentz-invariant:
\begin{equation}
d p^{\mu} = \frac{q}{mc} \, F^{\mu}_{\nu} \, p^{\nu},
\label{Lorentzforce}
\end{equation}
where $p^{\mu} = [p^0, {\bf p} ]$ is the four-momentum and $F^{\mu}_{\; \; \nu} $ is the Maxwell-Faraday
tensor. Reliable experimental confirmations of the validity of Lorentz transformations concern measurable
quantities such as charges, currents, energies (frequencies) and momenta (wave vectors) much more than
the less intrinsic quantities which are the {\it differentials} of the space-time variables. In principle,
the Lorentz transformations could have been established by very precise observations of the Doppler effect alone.

While analyzing fundamental laws relating between them very different entities we establish
by the same token the dimensional factors enabling the comparison of phenomena coming from
different physical realms. Let us consider the most ancient relation of this type, Newton's
second law of mechanics:
\begin{equation}
{\bf a} = \frac{1}{m} \, {\bf F}.
\label{Newton2}
\end{equation}
We wrote this equation in a bit unusual way in order to make it conformal with other equations
of the same type, which will follow. In this form, it is easier to ask two important questions
concerning this equation:
\vskip 0.2cm
\indent
\hskip 0.4cm
a) Which side of this equation represents a {\it measurable} quantity, directly accessible
to classical devices such as measuring rods and clocks ?
\vskip 0.2cm
\indent
\hskip 0.4cm
b) Which side of the above equation represents the {\it cause}, and which describes the {\it effect}
resulting from that cause ?
\vskip 0.2cm
\indent
In the case of the above equation (\ref{Newton2}) the answer is quite obvious, since we know that before
 it was postulated by Newton, Galileo has determined gravitational acceleration in a series of brilliant
 experiments. Later on, Newton introduced the notion of {\it gravitational force} to acknowledge the cause 
of the acceleration, and the mass as the proportionality factor inherent to a given body.

It turned out soon that the force {\bf F} may symbolize the action of quite different physical phenomena like
 gravitation, electricity or inertia, and is not a primary cause, but rather a manner of intermediate bookkeeping.
The more realistic causes of acceleration - or rather of the variation of energy and momenta - are the intensities
of electric, magnetic or gravitational fields.
A similar analysis can be applied to the basic equations of General Relativity. The intrinsic acceleration caused by
gravity becomes the {\it geodesic deviation} visible and measurable as the {\it tidal effect}:
\begin{equation}
\frac{ d^2 \delta x^{\mu}}{ds^2} = R^{\mu}_{\; \; \nu \lambda \rho} \frac{ d x^{\nu}}{ds} \frac{ d x^{\lambda}}{ds} \, \delta x^{\rho},
\label{geodevi}
\end{equation}
where $\delta x^{\lambda}$ is the infinitesimal deviation between two free-falling objects, and $R^{\mu}_{\; \; \nu \lambda \rho} $ is the
Riemann tensor. 

Here again, the measurable quantity is on the left, while the cause provoking the deviation phenomenon (i.e. the acceleration)
is encoded in the components of the energy-momentum tensor on the right. The real cause being the source in the form of the energy-momentum
tensor on the right-hand side of Einstein's equations:
\begin{equation}
R_{\mu \nu} - \frac{1}{2} g_{\mu \nu} R = - \frac{8 \pi G}{c^3} T_{\mu \nu}.
\label{Einstein2}
\end{equation}
The hypothesis concerning the presence of dark matter in Galaxies results from the analysis of
star motions, which combined with the deviation equation (\ref{Einstein2}) lead to the conclusion
 that the energy-momentum tensor on the right-hand side must represent more masses than what
can be estimated from the visible distribution of matter.

Our questioning about the cause of measurable effects should not stop at the stage of {\it forces},
which are but expressions of effects of countless fundamental interactions, just like the thermodynamical
pressure is in fact an averaged result of countless atomic collisions. On a classical level, when
theory permits, the symbolical force can be replaced by a more explicit expression in which fields
responsible for the forces do appear, like in the case of the Lorentz force (\ref{Lorentzforce}).

But the fields acting on a test particle are usually generated by more or less distant charges and currents,
according to the formula giving the retarded four-potential $A_{\mu} (x^{\lambda})$:
\begin{equation}
A_{\mu} ({\bf r}, t) = \frac{1}{4 \pi c} \; \int \int \int \frac{j_{\mu} ({\bf r}', t- \frac{\mid {\bf r} 
- {\bf r}' \mid}{c})}{\mid {\bf r} - {\bf r}' \mid } 
d^3 {\bf r}'.
\label{retarded}
\end{equation}
then we get the field tensor given by
$$F_{\mu \nu} = \partial_{\mu} A_{\nu} - \partial_{\nu} A_{\mu}.$$
The macroscopic currents are generated by electrons' collective motion. A single electron whose wave function is a bi-spinor
gives rise to the Dirac current
\begin{equation}
j^{\mu} = {\psi}^{\dagger} \gamma^{\mu} \psi,
\label{Diraccurrent}
\end{equation}
with 
$${\psi}^{\dagger} = {\bar{\psi}}^T \gamma^5, \; \; \; {\rm where} \; \; \; \gamma^5 = \gamma^0 \gamma^1 \gamma^2 \gamma^3 = 
{\begin{pmatrix} {I_2 & 0 \cr 0 &  - I_2} \end{pmatrix}}, \; \; \; I_2 = {\begin{pmatrix} {1 & 0 \cr 0 &  1} \end{pmatrix}}.$$
In fact, the four-component complex function $\psi$ is composed of two two-component spinors, $\xi_{\alpha}$ and $\chi_{{\dot{\beta}}}$ 
\cite{Streater},
$$ \psi = {\begin{pmatrix}{\xi \cr \chi} \end{pmatrix}},$$
which are supposed to transform under two non-equivalent representations of the $SL(2, {\bf C})$ group:
\begin{equation}
\xi_{\alpha'} = S^{\alpha}_{\alpha'} \xi_{\alpha}, \; \; \; \chi_{{\dot{\beta}}'} = S^{{\dot{\beta}}}_{{\dot{\beta}}'} \chi_{{\dot{\beta}}'},
\label{xichitransform}
\end{equation}
The electric charge conservation is equivalent to the annulation of the four-divergence of $j^{\mu}$:
\begin{equation}
\partial_{\mu} j^{\mu} = \left( \partial_{\mu} {\psi}^{\dagger} \gamma^{\mu} \right) \psi +
{\psi}^{\dagger} \left( \gamma^{\mu} \partial_{\mu} \psi \right) = 0,\end{equation}
from which we infer that this condition will be satisfied if we have
\begin{equation}
\partial_{\mu} {\psi}^{\dagger} \gamma^{\mu} = - m {\psi}^{\dagger} \; \; {\rm and} \; \; \gamma^{\mu} \partial_{\mu} \psi = m \psi,
\label{Diracfromj}
\end{equation}
which is the Dirac equation. In terms of the spinorial components $\xi$ and $\chi$ the Dirac equation can be seen as a pair
of two coupled equations which can be written in terms of Pauli's $\sigma$-matrices:
$$ \left( - i \hbar \frac{1}{c} \frac{\partial \;}{\partial t} + mc \right) \xi = i \hbar {\bf \sigma} \cdot {\bf \nabla} \chi,$$
\begin{equation}
 \left( - i \hbar \frac{1}{c} \frac{\partial \;}{\partial t} - mc \right) \chi = i \hbar {\bf \sigma} \cdot {\bf \nabla} \xi.
\label{xichiequations}
\end{equation}
The relativistic invariance imposed on this equation is usually presented as follows: under a Lorentz transformation $\Lambda$
the $4$-current $j^{\mu}$ undergoes the following change:
\begin{equation}
j^{\mu} \rightarrow j^{\mu'} = \Lambda^{\mu'}_{\mu} j^{\mu}.
\label{jmutransf}
\end{equation}
This means that the matrices $\gamma^{\mu}$ must transform as components of a $4$-vector, too. Parallelly, the components of the bi-spinor
$\psi$ must be transformed in a way such as to leave the form of the equations (\ref{Diracfromj}) unchanged: writing symbolically
the transformation of $\mid \psi >$ as $\mid \psi' > = S \mid \psi >$, and $< \psi' \mid = < \psi \mid S^{-1}$, we
should have
\begin{equation}
j^{\mu'} = < \psi' \mid \gamma^{\mu'} \mid \psi' > = 
< \psi \mid S^{-1} \gamma^{\mu'}  S \mid \psi >  = \Lambda^{\mu'}_{\mu} j^{\mu} = 
\Lambda^{\mu'}_{\mu}  < \psi \mid  \gamma^{\mu} \mid \psi >
\label{gammapsitransform}
\end{equation}
from which we infer the transformation rules for gamma-matrices:
\begin{equation}
S^{-1} \gamma^{\mu'} S = \Lambda^{\mu'}_{\mu} \gamma^{\mu}.
\end{equation}

%The relativistic quantum mechanics combines the electron and positron states in a single Dirac bi-spinor $\psi$ comprising the
%two aforementioned two-component spinors, and transforming under a $4 \times 4$ representation of $SL(2, {\bf C})$ group.
The usual way of presenting the joint effect of a Lorentz transformation $\Lambda$ on the coordinates and the
wave function is as follows:
\begin{equation}
x^{\mu} \rightarrow x^{\mu'} = \Lambda^{\mu'}_{\nu} x^{\nu}, \; \; \; 
\psi' (x^{\mu'}) = \psi' (\Lambda^{\mu'}_{\nu} x^{\nu}) = S (\Lambda) \psi (x^{\nu}).
\label{psitransform}
\end{equation}
This formula suggests that the transformation $S$ of the states in the Hilbert space is imposed
by the Lorentz transformation acting on the space-time coordinates, but in fact, decrypted from
transformation properties of classical macro-objects such as wave vectors or the 4-momentum.

In view of the analysis of the causal chain, it seems more appropriate to write the same transformations
with $\Lambda$ depending on $S$:
\begin{equation}
\psi' (x^{\mu'}) = \psi' (\Lambda^{\mu'}_{\nu} (S) x^{\nu}) = S \psi (x^{\nu})
\label{lambdaes}
\end{equation}
This form of the same relation suggests that the transition from one quantum state to another,
represented by the unitary transformation $S$ is the primary cause that implies the transformation
of observed quantities such as the electric $4$-current, and as a final consequence, the apparent
transformations of time and space intervals measured with classical physical devices.

Although mathematically the two formulations are equivalent, it seems more plausible that the Lorentz
group resulting from the averaging of the action of the $SL(2, {\bf C})$ in the Hilbert space of states
contains less information than the original double-valued representation which is a consequence of the
particle-anti-particle symmetry, than the other way round.

In what follows, we shall draw physical consequences from this approach, concerning the strong interactions
in the first place.

\section{Pauli's exclusion principle and the $SL(2, {\bf C})$ group}

The Pauli exclusion principle \cite{Pauli1}, according to which two electrons cannot be in the same state
characterized by identical quantum numbers, is one of the most important cornerstones
of quantum physics. This principle not only explains the structure of atoms and
therefore the entire content of the periodic table of elements, but it also guarantees 
the stability of matter preventing its collapse, as suggested by Ehrenfest \cite{Dyson1},
and proved later by Dyson \cite{Dyson2}, \cite{Dyson3}. The relationship between the
exclusion principle and particle's spin, known under the name of the ``spin-and-statistic theorem",
represents one of the deepest results in quantum field theory.

In purely algebraical terms Pauli's exclusion principle amounts to the anti-symmetry of 
wave functions describing two coexisting particle states. The easiest way to see how the principle
works is to apply Dirac's formalism in which wave functions of particles in given state are
obtained as products between the ``bra" and ``ket" vectors \cite{Dirac}.

Consider the probability amplitude to find a particle in the state $\mid x >$, 
\begin{equation}
\Phi (x) = < \psi \mid x>.
\label{x-state}
\end{equation}

The wave function of a two-particle state of which one is in the state $\mid x>$ and another
in the state $\mid y >$ is represented by a superposition
\begin{equation}
\mid \psi > = \sum \, \Phi (x,y) \, ( \mid x> \otimes \mid y>).
\label{xy-state}
\end{equation}
It is clear that if the wave function $\Phi (x,y)$ is anti-symmetric, i.e. if it satisfies
\begin{equation}
\Phi (x,y) = - \Phi (y,x),
\label{antisymPhi}
\end{equation}
then $\Phi (x,x) = 0$ and such states have vanishing both their wave function and probability.
It is easy to prove using the superposition principle, that this condition is not only
sufficient, but also necessary.
Let us suppose that $\Phi (x,x)$ vanish. This should remain valid in any basis  provided
the new basis $\mid x'>, \; \mid y'>$ was obtained from the former one via an unitary transformation.
Let us form an arbitrary state being a linear combination of $\mid x>$ and $\mid y>$,
$$ \mid z> = \alpha \mid x > + \beta \mid y >, \; \; \; \alpha, \beta \in {\bf C},$$
 and let us form the wave function of a tensor product of such a state with itself:
\begin{equation}
\Phi (z,z) = < \psi \mid ( \alpha \mid x > + \beta \mid y > ) \otimes ( \alpha \mid x > + \beta \mid y > ),
\label{Atensor}
\end{equation}
which develops as follows:
$$ \alpha^2 \, < \psi \mid (x,x)> + \alpha \beta  < \psi \mid (x,y) > + \beta \alpha \, < \psi \mid (y,x) > + 
\beta^2 \,  < \psi \mid (y,y) > =$$
\begin{equation}
= \Phi (x,y) = \alpha^2 \,  \Phi (x,x) + \alpha \beta \,  \Phi (x,y) 
+ \beta \alpha \, \Phi (y,x) + \beta^2 \, \Phi (y,y).
\label{Adeveloped}
\end{equation}
Now, as $ \Phi (x,x) = 0$ and $\Phi (y,y)=0$, the sum of remaining two terms will vanish if and only if (\ref{antisymPhi}) is
 satisfied, i.e. if $\Phi (x,y)$ is anti-symmetric in its two arguments.

After second quantization, when the states are obtained with creation and annihilation operators acting
on the vacuum, the anti-symmetry is encoded in the anti-commutation relations
%\begin{equation}
%a^{\dagger} (x) a^{\dagger} (y) + a^{\dagger} (y) a^{\dagger} (x) = 0.
%\label{anticommutation}
%\end{equation}

\begin{equation}
\psi (x) \psi(y) + \psi (y) \psi (x) = 0 \; \; \; 
\label{anticompsi}
\end{equation}
where $\psi(x) \mid 0> = \mid \psi >$.

Now, according to the experiment, electrons having identical energy and momenta can still display two different
states; in fact, this is the only possibility for two electrons to occupy otherwise identical states.
This is why for a given principal quantum number $n$ there are only $2 n^2$ possible electron states in
the corresponding electron shell. Therefore,
if these states (which are just two opposite directions of spin) are labeled $\mid 1>$ and $\mid 2>$, their
tensor product should contain only the anti-symmetric sector,
$$\mid 1> \otimes \mid 2> = - \mid 2> \otimes \mid 1>, $$
This property can be also expressed by admitting the existence of an anti-symmetric two-form in the
Hilbert space of two-electron states, which can be normalized to $1$ as follows:
$$\epsilon_{\alpha \beta} = - \epsilon_{\beta \alpha}, \; \; \alpha, \beta = 1,2; \; \; \; 
\epsilon^{12} = - \epsilon^{21} = 1, \; \; \; \epsilon^{11} = 0, \; \; \epsilon^{22} = 0.$$
According to the superposition principle, another basis in the Hilbert space of two-electron states
can be chosen; however, the Pauli principle should hold independently of such transformation. 
One should have then, after a linear transformation
$$\mid \psi_{\alpha} > \rightarrow \mid \psi_{\alpha'} > =  S^{\alpha}_{\alpha'} \; \mid \psi_{\alpha} >,$$
the same anti-symmetric $2$-form:
\begin{equation}
\epsilon_{\alpha' \beta'} = S^{\alpha}_{\alpha'} \; S^{\beta}_{\beta'} \;\epsilon_{\alpha \beta} 
\label{epsilontransform}
\end{equation}
Requiring the invariance of the form $\epsilon^{\alpha \beta}$, i.e. postulating that $\epsilon^{\alpha' \beta'}$
has the same components as before,
$$\epsilon_{1' 2'} = - \epsilon_{2' 1'} = 1, \; \; \; \epsilon_{1' 1'} = 0, \; \; \; \epsilon_{2' 2'} = 0$$
leads to the unique condition on the components of the complex $2 \times 2$ matrix $S$, namely
\begin{equation}
{\rm det} S = 1,
\label{det1}
\end{equation}
which determines the group $SL(2, {\bf C})$ as the invariance group of tensor products of electron states.

The existence of the positron implies the existence of a different, although similar sector of Hilbert space of states.
The complex conjugate matrices of the $SL(2, {\bf C})$ group yield another representation, which is not equivalent.
It acts on the complex conjugate spinors, whose indeces are labeled with dots, to make the difference clearly visible
(\cite{Streater}). We have therefore a skew-symmetric two-form
$$\epsilon_{{\dot{\alpha}} {\dot{ \beta}}} = - \epsilon_{{\dot{\beta}} {\dot{\alpha}}}, 
\; \; \; {\dot{\alpha}}, {\dot{\beta}} = 1,2; \; \; \; 
\epsilon^{{\dot{1}} {\dot{2}}} = - \epsilon^{{\dot{2}} {\dot{1}}} = 1, \; \; \; \epsilon^{{\dot{1}} {\dot{1}}} = 0, \; \; 
\epsilon^{{\dot{2}} {\dot{2}}} = 0.$$

A two-form with mixed indeces can be introduced, too: $x_{\alpha {\dot{\beta}}}$, transforming under the comined
action of $SL(2, {\bf C})$ and its conjugate representation ${\bar{SL}}(2, {\bf C})$. If we require it to be
hermitian, $x_{\alpha {\dot{\beta}}} = {\bar{x}}_{{\dot{\alpha}} \beta}$, its spans a four-dimensional linear
subspace whose basis can be labeled with an index $\mu$: $x^{\mu}_{\alpha {\dot{\beta}}}$, which transforms under
the $4 \times 4$ real representation of the Lorentz group \cite{Streater}.

\section{Ternary generalization of Pauli's principle}

The electrons and positrons satisfying Dirac's equation are considered as elementary particles 
not only because their propagation is well described by the solutions of this equation and
because they satisfy other physical predictions like the gyromagnetic ratio equal to $2$, but also 
because there is no experimental evidence of any internal structure.

The situation is quite different when protons and neutrons are being considered. Although at first
approximation their behavior can be also described quite successfully by the same Dirac equation, 
only with different mass parameter (and zero electric charge in the case of the neutron), their
physical parameters other than spin do not display values imposed by the Dirac equation. The magnetic
momentum is different and does not have the required gyromagnetic ratio. Moreover, high energy
experiments known as the {\it deep inelastic scattering} show that the gamma-photons with very high
energy penetrate inside the nucleon and are scattered by almost point-like entities, whose
characteristic dimensions must be at least three orders smaller than that of the proton itself:
about $10^{-16} cm$ versus $10^{-13}$ cm. These conctituents of nucleons are called {\it quarks},
and their characterstic dimensions are close to that of the electron. Apparently, some form of the
Lorentz symmetry is valid also for quarks; however, like in the case of the electrons, one
may think that it is not imposed by what is happening in the macroscopic world, but is a result
of the action of certain representation of $SL(2, {\bf C})$ in the Hilbert space of quantum states of quarks.

In the formalism called "Quantum Chromo-Dynamics" (QCD) quarks are considered as fermions, endowed  
with spin $\frac{1}{2}$. Free quarks are inaccessible for direct observation, only {\it three}
quarks or anti-quarks can coexist inside a fermionic baryon (respectively, anti-baryon), and a pair 
quark-antiquark can form a meson with integer spin. Besides, they have to belong to different
{\it colors}, also a three-valued set. There are two quarks in the first generation, 
$u$ and $d$ (``up" and ``down"), which may be considered as two states of a more general object,
just like proton and neutron are regarded upon as two isospin component of a doublet called
``nucleon". With this in mind we see that in the same bound state there is place for {\it two}
quarks in the same $u$-state or $d$-state, but not three.

This suggests that a convenient generalization of Pauli's exclusion principle would be the
statement that no {\it three} quarks in the same state can be present in a nucleon. Let us
require then the vanishing of wave functions corresponding to the tensor product of {\it three}
(but not necessarily two) identical states. That is, we require that $\Phi (x,x,x) = 0$ for
any state $\mid x>$. As in the former case, it is easy to prove that the necessary symmetry
condition for that to be true in any basis is
\begin{equation}
\Phi (x,y,z) + \Phi(y,z,x) + \Phi (z,x,y) = 0.
\label{threephi}
\end{equation}
Let us consider an arbitrary superposition of three different
states, $\mid x>, \; \mid y>$ and $\mid z>$,
$$ \mid w> = \alpha \mid x> + \beta \mid y> + \gamma \mid z>$$
and apply the same criterion, $\Phi (w,w,w) = 0$. We get then, after developing the tensor
products,
$$\Phi (w,w,w) = \alpha^3 \Phi(x,x,x) + \beta^3 \Phi( y,y,y) + \gamma^3 \Phi (z,z,z) $$
$$+ \alpha^2 \beta [ \Phi(x,x,y) + \Phi(x,y,x) + \Phi(y,x,x)] +
\gamma \alpha^2  [\Phi(x,x,z) + \Phi(x,z,x) + \Phi(z,x,x)] $$
$$+ \alpha \beta^2  [\Phi(y,y,x) + \Phi(y,x,y) + \Phi(x,y,y)] 
+ \beta^2 \gamma [\Phi(y,y,z) + \Phi(y,z,y) + \Phi(z,y,y)]$$
$$+ \beta \gamma^2 [\Phi(y,z,z) + \Phi(z,z,y) + \Phi(z,y,z)] 
+ \gamma^2 \alpha [\Phi(z,z,x) + \Phi(z,x,z) + \Phi(x,z,z)] $$
\begin{equation}
+ \alpha \beta \gamma [ \Phi(x,y,z) + \Phi(y,z,x) + \Phi(z,x,y) + \Phi(z,y,x)
+ \Phi(y,x,z) + \Phi(x,z,y) ]=0.
\label{fullthreeperm}
\end{equation}

%$$ < \psi \mid ( \alpha \mid x > + \beta \mid y > + \gamma \mid z> ) \otimes 
%( \alpha \mid x> + \beta \mid y + \gamma \mid z> ) \otimes ( \alpha \mid x > + \beta \mid y >
%+ \gamma \mid z> )=$$

The three diagonal expressions $\Phi (x,x,x), \; \Phi (y,y,y)$ and $\Phi (z,z,z)$ vanish by 
virtue of the original assumption; in what remains, every combination preceded by an independent
powers of three independent numerical coefficients $\alpha, \beta $ and $\gamma$, must vanish
separately. 

This can be achieved if the following $Z_3$ symmetry is imposed on the wave functions
of three arguments:
\begin{equation}
\Phi(x,y,z) = j \, \Phi (y,z, x) = j^2 \, \Phi (z,x,y), \; \; \; \; {\rm with} \; \; 
j=e^{\frac{2 \pi i}{3}}
\label{constituent1}
\end{equation}
where $j=e^{\frac{2 \pi i}{3}}$ is the cubic root of unity, satisfying $j^3 = 1, \; \; j+j^2 +1 =0$
Note that the complex conjugates of functions $\Psi (x,y,z)$ transform under cyclic permutations
of their arguments with $j^2 = \bar{j}$ replacing $j$ in the formula (\ref{constituent1}):
\begin{equation}
\Phi(x,y,z) = j^2 \, \Phi (y,z,x) = j \, \Phi (z,x,y).
\label{constituent2}
\end{equation}

In terms of operators acting on vacuum state producing states with definite number of quarks
(or antiquarks) this property will be translated into the following {\it cubic commutation relation}
generalizing Pauli's principle in the $Z_3$-graded case:
\begin{equation}
\theta^A \theta^B \theta^C = j \, \theta^B \theta^C \theta^A = j^2 \, \theta^C \theta^A \theta^B,
\label{ternary1}
\end{equation}

with $j = e^{2 i \pi/3}$, the primitive root of $1$. We have $1+j+j^2 = 0$
 and ${\bar{j}} = j^2$.

We shall also introduce a similar set of {\it conjugate} generators, 
 ${\bar{\theta}}^{\dot{A}}$,
{\small $\dot{A}, \dot{B},... = 1,2,...,N$,} satisfying similar condition with 
$j^2$ replacing  $j$:

\begin{equation}
{\bar{\theta}}^{\dot{A}} {\bar{\theta}}^{\dot{B}} {\bar{\theta}}^{\dot{C}} = 
j^2 \, {\bar{\theta}}^{\dot{B}} {\bar{\theta}}^{\dot{C}} {\bar{\theta}}^{\dot{A}} 
= j \, {\bar{\theta}}^{\dot{C}} {\bar{\theta}}^{\dot{A}} {\bar{\theta}}^{\dot{B}},
\label{ternary2}
\end{equation}
The two sets can be endowed with a natural $Z_3$ grading, with complex numbers being of grade $0$,
the generators $\theta^A$ being of grade $1$, their binary products of grade $2$, and ternary products of grade $0$ again, 
because the $Z_3$ grades add up modulo $3$. The conjugate generators are of grade $2$, their binary products of grade
$2+2 = 4 =1$ (modulo $3$), and the cubic products of grade $0$.

A direct consequence of these constitutive relations is the impossibility of forming products
of more than three quark or anti-quark operators. The proof is straightforward, using the associativity:
$$\theta^A \theta^B \theta^C \theta^D = j \, \theta^B \theta^C \theta^A \theta^D =
j^2 \, \theta^B \theta^A \theta^D \theta^C  =j^3 \, \theta^A \theta^D \theta^B \theta^C =
j^4 \, \theta^A \theta^B \theta^C \theta^D,$$
and because $j^4 = j \neq 1$, the only solution is $\theta^A \theta^B \theta^C \theta^D = 0.$
The same is true for the conjugate operators ${\bar{\theta}}^{\dot{A}}$.

In order to make the constitutive relations for the set of operators $\theta^{A}$ and ${\bar{\theta}}^{\dot{B}}$
complete, we have to impose {\it binary} commutation relations between the ``quark" and ``anti-quark" generators.
Looking at the cubic commutation relations (\ref{ternary1}) and (\ref{ternary2}) we see that a product of
two generators $\theta^A \theta^B$ behaves as an element of grade $2$;
$$(\theta^A \theta^B) \theta^C) = j^2 \, \theta^C (\theta^A \theta^B)$$
However, in order to distinguish between a product of two $\theta$'s and a conjugate generator ${\bar{\theta}}$
we shall impose an alternative choice possible for binary products:
\begin{equation}
\theta^A {\bar{\theta}}^{\dot{B}} = - j {\bar{\theta}}^{\dot{B}} \theta^A, \; \; \; 
{\bar{\theta}}^{\dot{B}} \theta^A = - j^2 \theta^A {\bar{\theta}}^{\dot{B}}.
\label{commutation2}
\end{equation}

\section{The invariance groups}

The constitutive cubic relations between the generators of the $Z_3$ graded algebra
can be considered as intrinsic if they are conserved after linear transformations with commuting 
(pure number) coefficients, i.e. if they are independent of the choice of the basis.

Let $U^{A'}_A$ denote a non-singular $N \times N$ matrix, transforming the generators
$\theta^A$ into another set of generators, $\theta^{B'} = U^{B'}_B \, \theta^B$.

We are looking for the solution of the covariance condition for the $\rho$-matrices:
\begin{equation}
\Lambda^{{\alpha}'}_{\beta} \, \rho^{{\beta}}_{ABC} = U^{A'}_{A} \, U^{B'}_B \, U^{C'}_C \, 
\rho^{{\alpha}'}_{A' B' C'}.
\label{covtrans1}
\end{equation}

Now, $\rho^{1}_{121} = 1$, and we have two equations corresponding to the choice of values of the index $\alpha'$
equal to $1$ or $2$. For $\alpha' = 1'$ the $\rho$-matrix on the right-hand side is $\rho^{1'}_{A' B' C'}$,
which has only three components,
$$\rho^{1'}_{1' 2' 1'}=1, \, \ \ \, \rho^{1'}_{2' 1' 1'}=j^2, \, \ \ \, \rho^{1'}_{1' 1' 2'}=j, $$
which leads to the following equation:

{\small
\begin{equation}
\Lambda^{1'}_{1} = U^{1'}_{1} \, U^{2'}_2 \, U^{1'}_1 + j^2 \, U^{2'}_{1} \, U^{1'}_2 \, U^{1'}_1 
+ j \, U^{1'}_{1} \, U^{1'}_2 \, U^{2'}_1 = U^{1'}_{1} \, (U^{2'}_2 \, U^{1'}_1 - U^{2'}_{1} \, U^{1'}_2),
\label{invariant1}
\end{equation} }
because $j^2 + j = - 1$.
For the alternative choice $\alpha' = 2'$ the $\rho$-matrix on the right-hand side is $\rho^{2'}_{A' B' C'}$,
whose three non-vanishing components are
$$\rho^{2'}_{2' 1' 2'}=1, \, \ \ \, \rho^{2'}_{1' 2' 2'}=j^2, \, \ \ \, \rho^{2'}_{2' 2' 1'}=j. $$
The corresponding equation becomes now:
{\small
\begin{equation}
\Lambda^{2'}_{1} = U^{2'}_{1} \, U^{1'}_2 \, U^{2'}_1 + j^2 \, U^{1'}_{1} \, U^{2'}_2 \, U^{2'}_1 
+ j \, U^{2'}_{1} \, U^{2'}_2 \, U^{1'}_1 = U^{2'}_{1} \, (U^{1'}_2 \, U^{2'}_1 - U^{1'}_{1} \, U^{2'}_2),
\label{invariant2}
\end{equation} }
The two remaining equations are obtained in a similar manner. We choose now the three lower indices
on the left-hand side equal to another independent combination, $(212)$. Then the $\rho$-matrix on the
left hand side must be $\rho^2$ whose component $\rho^2_{212}$ is equal to $1$. This leads to the
following equation when $\alpha' = 1'$:
{\small
\begin{equation}
\Lambda^{1'}_{2} = U^{1'}_{2} \, U^{2'}_1 \, U^{1'}_2 + j^2 \, U^{2'}_{2} \, U^{1'}_1 \, U^{1'}_2
+ j \, U^{1'}_{2} \, U^{1'}_1 \, U^{2'}_2 = U^{1'}_{2} \, (U^{1'}_2 \, U^{2'}_1 - U^{1'}_{1} \, U^{2'}_2),
\label{invariant3}
\end{equation}
and the fourth equation corresponding to $\alpha' = 2'$ is:
\begin{equation}
\Lambda^{2'}_{2} = U^{2'}_{2} \, U^{1'}_1 \, U^{2'}_2 + j^2 \, U^{1'}_{2} \, U^{2'}_1 \, U^{2'}_2
+ j \, U^{2'}_{2} \, U^{2'}_1 \, U^{1'}_2 = U^{2'}_{2} \, (U^{1'}_1 \, U^{2'}_2 - U^{2'}_{1} \, U^{1'}_2).
\label{invariant4}
\end{equation}}
%{The invariance group of cubic matrices}
%
%The determinant of the $2 \times 2$ complex matrix $U^{A'}_B$ appears everywhere on the right-hand side.
%The obvious solution relating {\it linearly} the matrices $\Lambda^{\alpha'}_{\beta}$ to the matrices
%$U^{A'}_B$ is to impose 
%\begin{equation}
%det \, (U^{A'}_B ) = U^{1'}_1 \, U^{2'}_2 - U^{2'}_{1} \, U^{1'}_2 = 1
%\label{detU}
%\end{equation}
%Then we have
%\begin{equation}
%\Lambda^{1'}_{1} = U^{1'}_1, \, \ \, \Lambda^{2'}_{2} = U^{2'}_2, \, \ \ 
%\Lambda^{1'}_{2} = - U^{1'}_2, \, \ \, \Lambda^{2'}_{1} = - U^{2'}_1,
%\label{lambdaUmatrices}
%\end{equation}
%from which it follows immediately that also $det \, \Lambda = 1$,
%\begin{equation}
%det \, (\Lambda^{\alpha'}_{\beta} ) = \Lambda^{1'}_1 \, \Lambda^{2'}_2 - \Lambda^{2'}_{1} \, \Lambda^{1'}_2 = 1
%\label{detlambda}
%\end{equation}
%Both conditions (\ref{detU}) and (\ref{detlambda}) define the $SL(2, {\bf C})$ group, the covering group
%of the Lorentz group.
%The determinant of the $2 \times 2$ complex matrix $U^{A'}_B$ appears everywhere on the right-hand side.
\begin{equation}
\Lambda^{2'}_{1} = - U^{2'}_{1} \, [det(U)],
\label{invariantdet}
\end{equation} 
The remaining two equations are obtained in a similar manner, resulting in the following: 
\begin{equation}
\Lambda^{1'}_{2} = - U^{1'}_{2} \, [det(U)], \, \ \ \, \ \ \Lambda^{2'}_{2} = U^{2'}_{2} \, [det(U)].
\label{invariant34}
\end{equation}
The determinant of the $2 \times 2$ complex matrix $U^{A'}_B$ appears everywhere on the right-hand side.
Taking the determinant of the matrix $\Lambda^{{\alpha}'}_{\beta}$ one gets immediately
\begin{equation}
det \, ( \Lambda ) = [ det \, (U) ]^3.
\label{detLambdaU}
\end{equation}

%{The invariance group of cubic matrices}
%
%Taking into account that the inverse transformation should exist and have the same properties,
%we arrive at the conclusion that $det \, \Lambda = 1$,
%\begin{equation}
%det \, (\Lambda^{\alpha'}_{\beta} ) = \Lambda^{1'}_1 \, \Lambda^{2'}_2 - \Lambda^{2'}_{1} \, \Lambda^{1'}_2 = 1
%\label{detlambda}
%\end{equation}
%which defines the $SL(2, {\bf C})$ group, the covering group of the Lorentz group.
%

However, the $U$-matrices on the right-hand side are defined only up to the phase, which due to the
cubic character of the covariance relations 
%(\ref{invariant1} - \ref{invariant34}), 
and they can take on three different values:
$1$, $j$ or $j^2$, i.e. the matrices $j \, U^{A'}_B$ or $j^2 \,  U^{A'}_B$ satisfy the same relations
as the matrices $U^{A'}_B$ defined above. 
The determinant of $U$ can take on the values $1, \, j \,$ or $j^2$ if  $det (\Lambda) = 1$

%But for the time being, there we have no reason yet to impose the unitarity condition. It can be
%derived from the conditions imposed on the invariance of binary relations between $\theta^A$ and
%their conjugates ${\bar{\theta}}^{\dot{B}}$.

%{\bf The vector representation}

 A similar covariance requirement can be formulated with respect to the set of $2$-forms
mapping the quadratic quark-anti-quark combinations into a four-dimensional linear real space.
As we already saw, the symmetry (\ref{commutation2}) imposed on these expressions reduces their
number to four. Let us define two quadratic forms,  $\pi^{\mu}_{A {\dot{B}}}$
 and its conjugate ${\bar{\pi}}^{\mu}_{{\dot{B}} A}$
\begin{equation}
\pi^{\mu}_{A {\dot{B}}} \, \theta^A {\bar{\theta}}^{\dot{B}} 
\; \; \; {\rm and} \; \; \;  {\bar{\pi}}^{\mu}_{{\dot{B}} A}
\, {\bar{\theta}}^{\dot{B}} \theta^A.
\label{pisymmetry1}
\end{equation}
The Greek indices $\mu, \nu...$ take on four values, and we shall label them
$0,1,2,3$.

The four tensors $\pi^{\mu}_{A {\dot{B}}}$ and their hermitian conjugates 
${\bar{\pi}}^{\mu}_{{\dot{B}} A}$ define a bi-linear mapping from the product of
quark and anti-quark cubic algebras into a linear four-dimensional vector space, whose
structure is not yet defined.

Let us impose the following invariance condition:

\begin{equation} 
\pi^{\mu}_{A {\dot{B}}} \, \theta^A {\bar{\theta}}^{\dot{B}} = {\bar{\pi}}^{\mu}_{{\dot{B}} A}
{\bar{\theta}}^{\dot{B}} \theta^A.
\label{invbin}
\end{equation}

It follows immediately from (\ref{commutation2}) that
\begin{equation}
\pi^{\mu}_{A {\dot{B}}} = - j^2 \, {\bar{\pi}}^{\mu}_{{\dot{B}} A}.
\label{pisymmetry2}
\end{equation}
Such matrices are non-hermitian, and they can be realized by the following substitution:
\begin{equation}
\pi^{\mu}_{A {\dot{B}}} = j^2 \, i \, {\sigma}^{\mu}_{A {\dot{B}}}, \, \ \ \,
{\bar{\pi}}^{\mu}_{{\dot{B}} A} = - j \, i \, {\sigma}^{\mu}_{{\dot{B}} A}
\label{pidefinition2}
\end{equation}
where ${\sigma}^{\mu}_{A {\dot{B}}}$
are the unit $2 \time 2$ matrix for $\mu = 0$, and the three hermitian Pauli matrices for $\mu = 1,2,3$.

Again, we want to get the same form of these four matrices in another basis. Knowing
that the lower indices $A$ and ${\dot{B}}$ undergo the transformation with matrices $U^{A'}_B$
and ${\bar{U}}^{{\dot{A}}'}_{\dot{B}}$, we demand that there exist some $4 \times 4$ matrices
$\Lambda^{{\mu}'}_{\nu}$ representing the transformation of lower indices by the matrices
$U$ and ${\bar{U}}$:
 \begin{equation}
\Lambda^{{\mu}'}_{\nu} \, \pi^{\nu}_{A {\dot{B}}} = U^{A'}_A \, {\bar{U}}^{{\dot{B}}'}_{\dot{B}}
 \pi^{{\mu}'}_{A' {\dot{B}}'},
\label{pitransform1}
\end{equation}
It is clear that we can replace the matrices  $\pi^{\nu}_{A {\dot{B}}}$ by the corresponding
matrices $\sigma^{\nu}_{A {\dot{B}}}$,
and this defines the vector ($4 \times 4)$ representation of the Lorentz group.

%The first four equations relating the $4 \times 4$ real matrices 
% $\Lambda^{{\mu}'}_{\nu}$ with the $2 \times 2$ complex matrices
% $U^{A'}_B$  and 
% ${\bar U}^{{\dot{A}}'}_{\dot{B}}$  are as follows:
%
% $$\Lambda^{0'}_0 - \Lambda^{0'}_3 = U^{1'}_2 \, {\bar U}^{{\dot{1}}'}_{\dot{2}} 
%+ U^{2'}_2 \, {\bar U}^{{\dot{2}}'}_{\dot{2}} $$
% $$\Lambda^{0'}_0 + \Lambda^{0'}_3 = U^{1'}_1 \, {\bar U}^{{\dot{1}}'}_{\dot{1}} 
%+ U^{2'}_1 \, {\bar U}^{{\dot{2}}'}_{\dot{1}} $$
% $$\Lambda^{0'}_0 - i \Lambda^{0'}_2 = U^{1'}_1 \, {\bar U}^{{\dot{1}}'}_{\dot{2}} 
%+ U^{2'}_1 \, {\bar U}^{{\dot{2}}'}_{\dot{2}} $$
% $$\Lambda^{0'}_0 + i \Lambda^{0'}_2 = U^{1'}_2 \, {\bar U}^{{\dot{1}}'}_{\dot{1}} 
%+ U^{2'}_2 \, {\bar U}^{{\dot{2}}'}_{\dot{1}} $$
%
%The next four equations relating the $4 \times 4$ real matrices 
% $\Lambda^{{\mu}'}_{\nu}$ with the $2 \times 2$ complex matrices
% $U^{A'}_B$  and 
% ${\bar U}^{{\dot{A}}'}_{\dot{B}}$ are as follows:
%
% $$\Lambda^{1'}_0 - \Lambda^{1'}_3 = U^{1'}_2 \, {\bar U}^{{\dot{2}}'}_{\dot{2}} 
%+ U^{2'}_2 \, {\bar U}^{{\dot{1}}'}_{\dot{2}} $$
% $$\Lambda^{1'}_0 + \Lambda^{1'}_3 = U^{1'}_1 \, {\bar U}^{{\dot{2}}'}_{\dot{1}} 
%+ U^{2'}_1 \, {\bar U}^{{\dot{1}}'}_{\dot{1}} $$
% $$\Lambda^{1'}_1 - i \Lambda^{1'}_2 = U^{1'}_1 \, {\bar U}^{{\dot{2}}'}_{\dot{2}} 
%+ U^{2'}_1 \, {\bar U}^{{\dot{1}}'}_{\dot{2}} $$
% $$\Lambda^{1'}_1 + i \Lambda^{1'}_2 = U^{1'}_2 \, {\bar U}^{{\dot{2}}'}_{\dot{1}} 
%+ U^{2'}_2 \, {\bar U}^{{\dot{1}}'}_{\dot{1}} $$

It can be checked that now ${\rm det} \; 
(\Lambda) = \left[ {\rm det} U \right]^2 \, \left[ {\rm det} {\bar{U}} \right]^2.$

The group of transformations thus defined is $SL(2, {\bf C})$, which is the covering group of the Lorentz group. 

%{The metric tensor $g_{\mu \nu}$}

With the invariant ``spinorial metric" in two complex dimensions, $\varepsilon^{AB}$
and $\varepsilon^{{\dot{A}}{\dot{B}}}$ such that $\varepsilon^{12} = - \varepsilon^{21} = 1$
and $\varepsilon^{{\dot{1}}{\dot{2}}} = - \varepsilon^{{\dot{2}}{\dot{1}}}$, we can define
the contravariant components $\pi^{\nu \, \, A {\dot{B}}}$. It is easy to show that the
Minkowskian space-time metric, invariant under the Lorentz transformations, can be defined as
\begin{equation}
g^{\mu \nu} = \frac{1}{2} \biggl[ \pi^{\mu}_{A {\dot{B}}} \, \pi^{\nu \, \, A {\dot{B}}} \biggr] 
= diag (+,-,-,-)
\label{Mmetric}
\end{equation}
Together with the anti-commuting spinors ${\psi}^{\alpha}$ the four real coefficients defining
a Lorentz vector, $x_{\mu} \, {\pi}^{\mu}_{A {\dot{B}}}$, can generate now the supersymmetry
via standard definitions of super-derivations.
Let us then choose the matrices $\Lambda^{\alpha'}_{\beta}$ to be the usual spinor representation of
the $SL(2, {\bf C})$ group, while the matrices $U^{A'}_{B}$ will be defined as follows:
\begin{equation}
U^{1'}_{1} = j \Lambda^{1'}_1,   U^{1'}_{2} = - j \Lambda^{1'}_2, 
U^{2'}_{1} = - j  \Lambda^{2'}_1,  U^{2'}_{2} = j \Lambda^{2'}_2, 
\label{Umatrices}
\end{equation}
the determinant of $U$ being equal to $j^2$. 
Obviously, the same reasoning leads to the conjugate cubic representation of $SL(2, {\bf C})$ if we require
the covariance of the conjugate tensor 
$${\bar{\rho}}^{\dot{\beta}}_{{\dot{D}}{\dot{E}}{\dot{F}}} = j \, 
{\bar{\rho}}^{\dot{\beta}}_{{\dot{E}}{\dot{F}} {\dot{D}}}
= j^2 \, {\bar{\rho}}^{\dot{\beta}}_{{\dot{F}} {\dot{D}} {\dot{E}}},$$
by imposing the equation similar to (\ref{covtrans1})
\begin{equation}
\Lambda^{{\dot{\alpha}}'}_{{\dot{\beta}}} \, {\bar{\rho}}^{{\dot{\beta}}}_{{\dot{A}}{\dot{B}}{\dot{C}}} = 
{\bar{\rho}}^{{\dot{\alpha}}'}_{{\dot{A}}' {\dot{B}}' {\dot{C}}'} {\bar{U}}^{{\dot{A}}'}_{{\dot{A}}} \, 
{\bar{U}}^{{\dot{B}}'}_{{\dot{B}}} \, {\bar{U}}^{{\dot{C}}'}_{{\dot{C}}} .
\label{covtrans2}
\end{equation}
The matrix $\bar{U}$ is the complex conjugate of the matrix $U$, with determinant equal to $j$.

Moreover, the two-component entities obtained as images of cubic combinations of quarks,
$\psi^{\alpha} = \rho^{\alpha}_{ABC} \theta^A \theta^B \theta^C$ 
 and ${\bar{\psi}}^{\dot{\beta}} = {\bar{\rho}}^{{\dot{\beta}}}_{{\dot{D}}{\dot{E}}{\dot{F}}}
{\bar{\theta}}^{\dot{D}} {\bar{\theta}}^{\dot{E}} {\bar{\theta}}^{\dot{F}} $ should anti-commute,
because their arguments do so, by virtue of (\ref{commutation2}):
$$ (\theta^A \theta^B \theta^C) ({\bar{\theta}}^{\dot{D}} {\bar{\theta}}^{\dot{E}} {\bar{\theta}}^{\dot{F}} )
= - ({\bar{\theta}}^{\dot{D}} {\bar{\theta}}^{\dot{E}} {\bar{\theta}}^{\dot{F}})(\theta^A \theta^B \theta^C)$$
We have found the way to derive the covering group of the Lorentz group acting on spinors via
the usual spinorial representation. Spinors are obtained as a homomorphic image of 
tri-linear combinations of three quarks (or anti-quarks). The quarks transform with
matrices $U$ (or ${\bar{U}}$ for the anti-quarks), but these matrices are not unitary: 
their determinants are equal to $j^2$ or $j$, respectively. 

So, quarks cannot
be put on the same footing as classical spinors; they transform under the $Z_3 \times
SL(2, {\bf C})$ group. There are strong reasons to believe that their wave functions in the Schroedinger
picture should not obey exactly the same equations as the electrons; a modified version of
Dirac's equation should be found tào explain why they do not propagate as ordinary solutions do,
while their tri-linear combinations can propagate if extra selection rules (only combinations
with three different ``colors") display behavior similar to that of the ordinary spin-one-half
particles.

\section{A $Z_3$ generalization of Dirac's equation}

Lets us first underline the $Z_2$ symmetry of Maxwell and Dirac equations, which implies the hyperbolic
character of both systems, and therefore makes the propagation possible. Maxwell's equations {\it in vacuo}
can be written as follows:
$$\frac{1}{c} \frac{\partial {\bf E}}{\partial t} = {\bf \nabla} \wedge {\bf B},$$
\begin{equation}
- \frac{1}{c} \frac{\partial {\bf B}}{\partial t} = {\bf \nabla} \wedge {\bf E}.
\label{Maxwell2}
\end{equation}
These equations can be decoupled by applying the time derivation twice, which in vacuum, where
$div {\bf E} = 0$ (and $div {\bf B} =0$ which holds always) leads to the d'Alembert
equation satisfied by both components separately:
$$\frac{1}{c^2} \frac{\partial^2 {\bf E}}{\partial t^2} - {\bf \nabla}^2  {\bf E} = 0, \; \; \; \; \; \; 
\frac{1}{c^2} \frac{\partial^2 {\bf B}}{\partial t^2} - {\bf \nabla}^2 {\bf B} = 0.$$
Nevertheless, neither of the components of the Maxwell tensor, be it ${\bf E}$ or ${\bf B}$, can propagate
separately alone. It is also remarkable that although each of the firlds ${\bf E}$ and ${\bf B}$ satisfies
a second-order propagation equation, due to the coupled system (\ref{Maxwell2}) there exists a quadratic
combination satisfying the forst-order equation, the Poynting four-vector:
\begin{equation}
P^{\mu} = \left[ P^0, {\bf P} \right], \; \; \; 
P^0 = \frac{1}{2} \left( {\bf E}^2 + {\bf B}^2 \right), \; \; \; {\bf P} = {\bf E} \wedge {\bf B}, 
\; \; \; \partial_{\mu} P^{\mu} = 0.
\label{Poynting}
\end{equation}
The Dirac equation for the electron displays a similar $Z_2$ symmetry, with two coupled equations
which can be put in the following form:
$$i \hbar \frac{\partial }{\partial t} \psi_{+} - mc^2 \psi_{+} = 
 i \hbar {\bf \sigma} \cdot {\bf \nabla} \psi_{-},$$
\begin{equation}
- i \hbar \frac{\partial }{\partial t} \psi_{-} - mc^2 \psi_{-} = 
- i \hbar {\bf \sigma} \cdot {\bf \nabla} \psi_{+},
\label{Diracpmcoupled}
\end{equation}
where $\psi_{+}$ and $\psi_{-}$ are the positive and negative energy components of the
Dirac equation; this is visible even better in the momentum representation:
$$\left[ E -mc^2 \right] \psi_{+} = c {\bf \sigma} \cdot {\bf p} \psi_{-}, $$
\begin{equation}
\left[ -E - mc^2 \right] \psi_{-} = - c {\bf \sigma} \cdot {\bf p} \psi_{+}.
\end{equation}
Note that the same effect (negative energy states) can be obtained by changing the direction
of time, and putting the minus sign in front of the time derivative, as suggested
by Feynman \cite{Feynman}.
Each of the components satisfies the Klein-Gordon equation, easily obtained by successive
application of two operators and diagonalization:
$$ \left[  \frac{1}{c^2} \frac{\partial^2}{\partial t^2} - 
{\bf \nabla}^2  - m^2 \right] \psi_{\pm} =0$$
 As in the case of the electromagnetic waves, neither of the components of this complex entity
can propagate by itself; only all the components can.

As it follows from the experiment, the two types of quarks, $u$ and $d$, cannot propagate freely,
but can form a freely propagating particle perceived as a fermion, but only under an extra
condition: they must belong to three {\it different} species called {\it colors}; short of this
they will not form a freely propagating entity.

Therefore, quarks should be described by {\it three} fields satisfying a set of coupled linear equations,
with the $Z_3$ symmetry playing a similar role as the $Z_2$ symmetry did in the case of Maxwell's and
Dirac's equations. Instead of the ``-" sign multiplying the time derivative, we should use the cubic root of unity $j$ 
and its complex conjugate $j^2$ according to the following scheme:
$$\frac{\partial }{\partial t} \mid \psi > = {\hat H}_{12} \mid \phi >,$$
$$ j \frac{\partial }{\partial t} \mid \phi > = {\hat H}_{23} \mid \chi >,$$
\begin{equation}
j^2 \frac{\partial }{\partial t} \mid \chi > = {\hat H}_{31} \mid \psi >,
\label{threeeqs1}
\end{equation}
At the moment, we do not specify the number of components in each state vector, nor the
character of the hamiltonian operators on the right-hand side; the three fields $\mid \psi >, \mid \phi >$ and $\mid \chi >$
should represent the three colors, none of which can propagate by itself. The quarks being endowed with mass, we can
suppose that one of the main terms in the hamiltonians is the mass operator ${\hat{m}}$; and let us suppose that the
remaining parts are the same in all three hamiltonians. This will lead to the following three equations:
$$\frac{\partial }{\partial t} \mid \psi >  - {\hat{m}} \mid \psi > = {\hat H} \mid \phi >,$$
$$ j \frac{\partial }{\partial t} \mid \phi > - {\hat{m}} \mid \phi > = {\hat H} \mid \chi >,$$
\begin{equation}
j^2 \frac{\partial }{\partial t} \mid \chi > - {\hat{m}} \mid \chi > = {\hat H} \mid \psi >,
\label{threeeqs2}
\end{equation}
Supposing that the mass operator commutes with time derivation, we easily proven by applying three times
the left-hand side operators, that each of the components satisfies the same common {\it third order} equation:
\begin{equation}
\left[ \frac{\partial^3}{\partial t^3} - {\hat{m}}^3 \right] \mid \psi > = {\hat{H}}^3  \mid \psi >.
\label{thirdorderpsi}
\end{equation}
The anti-quarks should satisfy a similar equation with the negative sign for the Hamiltonian operator. The fact
that there exist two types of quarks in each nucleon suggests that the state vectors $\mid \psi >, \mid \phi >$ and $\mid \chi >$
should have two components each. When combined together, the two postulates lead to the conclusion that we must have
three two-component functions and their three conjugates:
$${\begin{pmatrix}{ \psi_1 \cr \psi_2} \end{pmatrix}}, \;  {\begin{pmatrix}{{\bar{\psi}}_{\dot{1}} \cr {\bar{\psi}}_{\dot{2}}} \end{pmatrix}};
\; \; \; \; \; \; \; \;
{\begin{pmatrix}{ \varphi_1 \cr \varphi_2} \end{pmatrix}}, \;  {\begin{pmatrix}{{\bar{\varphi}}_{\dot{1}} \cr {\bar{\varphi}}_{\dot{2}}} \end{pmatrix}} ;
\; \; \; \; \; \; \; \;
{\begin{pmatrix}{ \chi_1 \cr \chi_2} \end{pmatrix}}, \; \; {\begin{pmatrix}{{\bar{\chi}}_{\dot{1}} \cr {\bar{\chi}}_{\dot{2}}} \end{pmatrix}},$$
which may represent three colors, two quark states (e.g. ``up" and ``down"), and two anti-quark states (with anti-colors, respectively).

Finally, in order to be able to implement the action of the $SL(2, {\bf C})$ group via its $2 \times 2$ matrix representation
defined in the previous section, we choose the Hamiltonian ${\hat{H}}$ equal to the operator ${\bf \sigma} \cdot {\bf \nabla}$, the same as
in the usual Dirac equation. The action of the $Z_3$ symmetry is represented by factors $j$ and $j^2$, while the $Z_2$ symmetry between
particles and anti-particles is represented by the ``-" sign on the right-hand side. 

The differential system that satisfies all these assumptions is as follows:
$$ \left( - i \hbar \frac{\partial}{\partial t} - mc^2 \right) \psi =   i \hbar c {\bf \sigma} \cdot {\bf \nabla} {\bar{\varphi}},$$
$$ \left( - j i \hbar \frac{\partial}{\partial t} - mc^2 \right) {\bar{\varphi}} = - i \hbar c {\bf \sigma} \cdot {\bf \nabla} \chi,$$
$$ \left( - j^2 i \hbar \frac{\partial}{\partial t} - mc^2 \right) \chi =  i \hbar c {\bf \sigma} \cdot {\bf \nabla} {\bar{\psi}},$$
$$ \left( - i \hbar \frac{\partial}{\partial t} - mc^2 \right) {\bar{\psi}} = - i \hbar c {\bf \sigma} \cdot {\bf \nabla} \varphi,$$
$$ \left( - j i \hbar \frac{\partial}{\partial t} - mc^2 \right) \varphi =  i \hbar c {\bf \sigma} \cdot {\bf \nabla} {\bar{\chi}},$$
\begin{equation}
 \left(- j^2 i \hbar \frac{\partial}{\partial t} - mc^2 \right){\bar{\chi}} = - i \hbar c {\bf \sigma} \cdot {\bf \nabla} \psi,
\label{sixeqs}
\end{equation}
Here we made a simplifying assumption that the mass operator is just proportional to the identity matrix.

The functions $\psi, \varphi$ and $\chi$ are related to their conjugates via the following third-order equations:
$$
\left[ i  \frac{\partial^3}{\partial t^3} - \frac{ m^3 c^6}{{\hbar}^3} \right] \psi = - i ({\bf \sigma} \cdot {\bf \nabla})^3 = 
- i {\bf \sigma} \cdot {\bf \nabla} (\Delta {\bar{\psi}}), $$ 
\begin{equation}
\left[ i \frac{\partial^3}{\partial t^3} - \frac{ m^3 c^6}{{\hbar}^3}  \right] {\bar{\psi}} =  i ({\bf \sigma} \cdot {\bf \nabla})^3 = 
 i {\bf \sigma} \cdot {\bf \nabla} (\Delta \psi). 
\label{thirddiffeq}
\end{equation}
and the same, of course, for the remaining wave functions $\varphi$ and $\chi$.

The overall $Z_2 \times Z_3$ symmetry can be grasped much better if we use the matrix notation, encoding the
system of linear equations (\ref{sixeqs}) as an operator acting on a single vector composed of all the components.
In order to make it closer to the more familiar form of Dirac's equation, we shall

By consecutive application of these operators we can separate the variables and find the common equation of sixth order
that is satisfied by each of the components:
\begin{equation}
- {\hbar}^6 \frac{\partial^6}{\partial t^6} \psi - m^6 c^{12} \psi = - {\hbar}^6 {\Delta}^3 \psi.
\label{sixthorder}
\end{equation}
Identifying quantum operators of energy nand the momentum,
$$ - i {\hbar} \frac{\partial}{\partial t} \rightarrow E, \; \; \; - i {\hbar} {\bf \nabla} \rightarrow {\bf p},$$
we can write (\ref{sixthorder}) simply as follows:
\begin{equation}
E^6 - m^6 c^{12} = \mid {\bf p} \mid^6 c^6.
\label{Ep_relation}
\end{equation}
This equation can be factorized showing how it was obtained by subsequent action of the operators of the system (\ref{sixeqs}):
$$
E^6 - m^6 c^{12} = (E^3 - m^3 c^6)(E^3 + m^3 c^6) =$$
\begin{equation}
 (E - mc^2)(jE - mc^2)(j^2 E - mc^2)(E + mc^2)(jE + mc^2)(j^2 E + mc^2) = \mid {\bf p} \mid^6 c^6.
\label{Epfactorized}
\end{equation}
The equation (\ref{sixthorder}) can be solved by separation of variables; the time-dependent
and the space-dependent factors have the same structure:
$$A_1 \,e^{\omega\,t} + A_2 \,e^{j \,\omega\,t} + A_3 e^{j^2 \,\omega\,t},\,
\ \ \ \ B_1\,e^{{\bf k.r}} + B_2\,e^{j\,{\bf k.r}} + B_3\,e^{j^2\,{\bf k.r}}$$
with $\omega$ and ${\bf k}$ satisfying the following dispersion relation:
\begin{equation}
\frac{\omega^6}{c^6} = \frac{m^6 c^6}{{\hbar}^6} + \mid {\bf k} \mid^6,
\label{dispersion6}
\end{equation}
where we have identified $E = {\hbar \omega}$ and ${\bf p} = {\hbar} {\bf k}$. This relation is invariant under
the action of $Z_2 \times Z_3$ symmetry, because to any solution with given real $\omega$ and ${\bf k}$ one can
add solutions with $\omega$ replaced by $j \omega$ or $j^2 \omega$, $j {\bf k}$ or $j^2 {\bf k}$, as well as 
$- \omega$; there is no need to introduce also $- {\bf k}$ instead of ${\bf k}$ because the vector
${\bf k}$ can take on all possible directions covering the unit sphere.

The nine complex solutions can be displayed in two $3 \times 3$ matrices as follows:
\begin{equation}
\begin{pmatrix} { e^{\omega\,t - {\bf k \cdot r}} & e^{\omega\,t - j {\bf k \cdot r}}
& e^{\omega\,t - j^2 {\bf k \cdot r}} \cr 
e^{j \omega\,t-{\bf k \cdot r }} & 
e^{j \omega\,t - j {\bf k \cdot r}} &
e^{j \omega\,t - j^2 {\bf k \cdot r}} \cr
e^{j^2 \omega\,t -{\bf k \cdot r}}& e^{j^2 \omega\,t - {\bf k \cdot r}}  & 
e^{j^2 \omega\,t - j^2 {\bf k \cdot r}} }
\end{pmatrix}, \; \; \; 
\begin{pmatrix} { e^{- \omega\,t - {\bf k \cdot r}} & e^{-\omega\,t - j {\bf k \cdot r}}
& e^{-\omega\,t - j^2 {\bf k \cdot r}} \cr 
e^{-j \omega\,t-{\bf k \cdot r }} & 
e^{-j \omega\,t - j {\bf k \cdot r}} &
e^{-j \omega\,t - j^2 {\bf k \cdot r}} \cr
e^{-j^2 \omega\,t -{\bf k \cdot r}}& e^{-j^2 \omega\,t - {\bf k \cdot r}}  & 
e^{-j^2 \omega\,t - j^2 {\bf k \cdot r}} }
\end{pmatrix}
\end{equation} 
and their nine independent  products can be represented in a basis of real
functions as
{\small
\begin{equation}
\begin{pmatrix} {A_{11} \, e^{\omega\,t - {\bf k \cdot r}} & A_{12} \, e^{\omega\,t + \frac{{\bf k \cdot r}}{2}}
\, \cos ({\bf k} \cdot {\bf \xi}) & A_{13} \, e^{\omega\,t + \frac{{\bf k \cdot r}}{2}} \, \sin ({\bf k} \cdot {\bf \xi}) \cr 
A_{21}\, e^{- \frac{\omega\,t}{2}-{\bf k \cdot r }} \, \cos \omega \tau & A_{22}\,
e^{- \frac{\omega\,t}{2}+ \frac{\bf k \cdot r}{2} } \, \cos (\omega \tau - {\bf k} \cdot {\bf \xi}) &
A_{23} \,e^{- \frac{\omega\,t}{2}+ \frac{{\bf k \cdot r}}{2}} \, \cos (\omega \tau  + {\bf k} \cdot {\bf \xi}) \cr
A_{31} \,e^{- \frac{\omega\,t}{2} -{\bf k \cdot r}} \, \sin \omega \tau & A_{32}\,
e^{- \frac{\omega\,t}{2}+ \frac{ {\bf k \cdot r}}{2}} \, \sin (\omega \tau + {\bf k} \cdot {\xi}) & A_{33}\,
e^{- \frac{\omega\,t}{2}+\frac{{\bf k \cdot r}}{2}} \, \sin (\omega \tau - {\bf k} \cdot {\bf \xi}) }
\end{pmatrix}
\label{bigmatrix}
\end{equation} }
where $\tau=\frac{\sqrt{3}}{2} \,t$ and  $\xi=\frac{\sqrt{3}}{2}{\bf kr}$; the same can be done with the conjugate solutions
(with $- \omega$ instead of $\omega$).

The functions displayed in the matrix do not represent a wave; however, one can produce a propagating
solution by forming certain cubic combinations, e.g. 
$$e^{\omega\,t - {\bf k \cdot r}} \, e^{- \frac{\omega\,t}{2} + \frac{{\bf k \cdot r}}{2}}\,cos (\omega \tau - {\bf k} \cdot {\bf \xi})  \,
e^{- \frac{\omega\,t}{2} + \frac{{\bf k \cdot r}}{2}}\,sin (\omega \tau - {\bf k} \cdot {\bf \xi} ) = \frac{1}{2} \,
 \sin ( 2  \omega \tau - 2 {\bf k} \cdot {\bf \xi}). $$

What we need now is a multiplication scheme that would define triple products of non-propagating solutions yielding
propagating ones, like in the example given above, but under the condition that the factors belong to three distinct
subsets b(which can be later on identified as ``colors"). This can be achieved with the $3 \times 3$ matrices
of three types, containing the solutions displayed in (\ref{bigmatrix}), distributed in a particular way, each of the three
matrices containing the elements of one particular line of matrix (\ref{bigmatrix}):
{\small
\begin{equation}
[A]= \begin{pmatrix} { 0 & A_{12} \, e^{\omega\,t - {\bf k \cdot r}} & 0 \cr 0 & 0 & A_{23} \, e^{\omega\,t + \frac{{\bf k \cdot r}}{2}}
\, \cos {\bf k} \cdot {\bf \xi} \cr 
A_{31} e^{\omega\,t + \frac{{\bf k \cdot r}}{2}} \, \sin {\bf k} \cdot {\bf \xi} & 0 & 0 }
\end{pmatrix}
\label{bigmatrixA}
\end{equation} }

{\small
\begin{equation}
[B]= \begin{pmatrix} { 0 & B_{12} \, e^{- \frac{\omega}{2}\,t - {\bf k \cdot r}}  \, \sin \tau & 0 \cr 
0 & 0 & B_{23} \, e^{- \frac{\omega}{2}\,t + \frac{{\bf k \cdot r}}{2}} \, \cos ( \tau + {\bf k} \cdot {\bf \xi}) \cr 
B_{31} e^{\omega\,t - {\bf k \cdot r}} \, \cos \tau  & 0 & 0 }
\end{pmatrix}
\label{bigmatrixB}
\end{equation} }

{\small
\begin{equation}
[C]= \begin{pmatrix} { 0 & C_{12} \, e^{-\frac{\omega}{2} \,t + \frac{1\bf k \cdot r}{2}} \, \sin (\tau + {\bf k} \cdot {\bf \xi})& 0 \cr 
0 & 0 & C_{23} \, e^{- \frac{\omega}{2} \,t + \frac{{\bf k \cdot r}}{2}} \, \sin (\tau  - {\bf k} \cdot {\bf \xi}) \cr 
C_{31} e^{\omega\,t + \frac{{\bf k \cdot r}}{2}} \, \sin {\bf k} \cdot {\bf \xi} & 0 & 0 }
\end{pmatrix}
\label{bigmatrixA}
\end{equation} }
This model can explain why a single quark cannot propagate, while three quarks can form a freely propagating state.

%\section{Preparing your paper}
%\verb"jpconf" requires \LaTeXe\ and  can be used with other package files such
%as those loading the AMS extension fonts 
%\verb"msam" and \verb"msbm" (these fonts provide the 
%blackboard bold alphabet and various extra maths symbols as well as 
%symbols useful in figure captions); an extra style file \verb"iopams.sty" is 
%provided to load these packages and provide extra definitions for bold Greek letters. 
%\subsection{Headers, footers and page numbers}
%Authors should {\it not} add headers, footers or page numbers to the pages of their article---they will
%be added by \iopp\ as part of the production process.

\end{document}